\documentclass[%
 reprint,
 amsmath,amssymb,
 aps,
]{revtex4-1}

\usepackage{graphicx}
\usepackage{dcolumn}
\usepackage{bm}
\usepackage[retainorgcmds]{IEEEtrantools}
\catcode`\"=13
\def"#1{\v #1}

\begin{document}

\preprint{APS/123-QED}

\title{Exact Generalization of Local Thermodynamics of Inhomogeneous
Fluids to Arbitrary Spatial Scales}

\author{Alja"z Godec}
\email{aljaz.godec@ki.si}
\author{Janko Jamnik}%
\author{Franci Merzel}%
\affiliation{%
 National Institute of Chemistry, Ljubljana, Slovenia
}%




\date{\today}

\begin{abstract}
In order to provide a formally correct thermodynamical description of
inhomogeneous fluids valid on all length scales down to the classical
limit we postulate that all extensive quantities have locally
extensive analogues. We derive local thermodynamic relations for  
the canonical and isothermal-isobaric ensembles without
limitations with respect to the nature of intermolecular
interactions. 
The inhomogeneity
of the system is shown to result in two distinct entropic
contributions. The {\it intrinsic} contribution is due the specific distribution of local
environments experienced by a single molecule while the {\it extrinsic} one originates exclusively from
the local variation of the density from its bulk value. As an example the theory is
applied to the surface tension of a spherical liquid droplet.

\end{abstract}

\maketitle
The study of spatially inhomogeneous fluids is important for a vast
variety of applications, ranging from macroscopic surface and interfacial
phenomena  \cite{Widom2}, spatially confined systems such
as fluids confined to channels of porous materials \cite{pores}, down
to molecular solvation phenomena \cite{solvation}. A
proper theoretical framework for addressing these problems should be
capable to unambiguously describe local thermodynamical properties.
The term 'local' is to be understood in terms of a
function that is well defined at a certain point in space and in general
also depends on the state of the system in the neighborhood of the
point. The size of the neighborhood as well as the spatial scale, on
which the description is valid, depend critically on the level of theory.   

Various
theoretical attempts to describe 'local' thermodynamic properties of
inhomogeneous fluids have been proposed (traditionally called {\it
  point thermodynamics}) relying either on
Kirkwoods integral equation \cite{Hill}, spatially-dependent
cluster expansions \cite{Stillinger} or a density functional treatment
\cite{Percus}. In the fist two approaches \cite{Hill,Stillinger} the definition of local
thermodynamic potentials necessitates postulating the existence of a
local pressure, which is justified on spatial scales, on
which local properties vary slowly, but it fails on smaller
scales, on which one has to introduce a pressure tensor due to spatial
inhomogeneity \cite{Percus}. A proper description which enables a
full coupling of local thermal and mechanical properties was made
using the density functional formalism \cite{Percus}. While it consistently
describes fluids near interfaces or confined in pores and also enables a
direct extension to the description of transport of dense
inhomogeneous fluids it builds on the
assumption that the correlation function is short ranged \cite{footnote}. Thus it cannot be applied to describe
systems on the scale of the correlation length or even below, 
such as of immediate interest, for example, in solvation phenomena \cite{solvation} or
molecules confined to nanotubes \cite{nanotubes}. In these systems one is often
specifically interested in entropic effects ({\it i.e.} the
constraining of molecular degrees of freedom). Inevitably one has to resort to some sort of correlation expansion
\cite{correlation}, which becomes impractical for
inhomogeneous systems and is therefore usually
{\it ad hoc} truncated at the pair correlation level. As
significant higher
order correlations were found already in homogeneous simple liquids
\cite{triplet} such a truncation can not be
appropriate in the case of hydration
\cite{hydration}. More importantly, the use of thermodynamic concepts in
the case when the dimension of the studied (sub)system is comparable
to or smaller than the correlation length, is formally not
justified. The immediate technological interest in such systems  
such as nanofluidic applications \cite{nanofluid} and
 wetting phenomena in nanotubes
\cite{nanotherm}, poses a pressing need for a general
formulation of local thermodynamics on such small length scales. At
present, such a formal framework is still elusive and even   
the question of mere validity of
thermodynamical concepts on spatial scales comparable to or smaller than the
correlation length is still under debate. 

Here we present an exact rigorous formulation of local thermodynamics of
inhomogeneous fluids at equilibrium, which holds
on arbitrary spatial scales down to the classical
limit without any limitations regarding intermolecular
interactions or the origin of the inhomogeneity. Based on the test
particle picture we show that the specification of the order of
correlations can be entirely avoided while they are inherently built
into the theory. Defining locally extensive properties we derive a
complete set of local thermodynamic relations in the canonical and
isothermal-isobaric ensemble. An exact expression for the surface
tension of a spherical liquid
droplet is derived   
as a 'proof of concept' example. Finally, we
discuss the physical meaning of local thermodynamic potentials on
scales smaller than the correlation length.

While the quantities pertaining to properties of single molecules
(e.g. local density, $\langle \rho (\bm{r}) \rangle$) have an
unambiguous local definition, functions which are averages
over the properties of larger groups of molecules (e.g. free energy density, $f(\bm{r})$, or entropy density, $s(\bm{r})$), cannot be unambiguously defined but are required to fulfill
certain constraints, typically the invariance of some of the lower
moments of the functions \cite{Widom2}. 
We postulate that all nominally extensive quantities have well defined local
analogues and therefore introduce the notion of a \textit{locally extensive}
quantity $a(\bm{r})$ as:
\begin{equation} 
a(\bm{r})  \equiv \langle \rho (\bm{r}) \rangle A^1(\bm{r}),
\label{eq1} 
\end{equation}
where the superscript $1$ denotes the property per single molecule
located in the volume element $\mathrm{d}\bm{r}$ centered in
$\bm{r}$. We describe the 'per molecule' or 'quasi-single' particle property
within the test particle-insertion picture, where a local quantity is
typically a function of the insertion energy of the
\textit{virtual} test molecule and assign the entire potential
energy change to the test molecule. We show that such a choice leads to invariant values of integrals of
local thermodynamic functions over
space. We focus on excess thermodynamic properties since the
corresponding ideal contributions are trivial to evaluate.  

In the canonical ensemble we begin by defining $f_{ex}(\bm{r})$ in terms of
the excess free
energy per molecule located in the differential volume element,
$\overline F _{ex}^1(\bm{r})$, which can be understood as the change in excess free energy
upon insertion of an additional molecule at position $\bm{r}$ at constant
$T$ and $V$ integrated over all orientations of the inserted molecule
and all configurations (positions as well as orientations) of the
remaining $N$ molecules. $\overline F _{ex}^1(\bm{r})$ is expressed   
in terms of the corresponding configurational integrals of the $(N+1)$- and $N$-particle
systems: $\overline F
_{ex}^1(\bm{r})=-k_BT\ln\frac{Z_{N+1}(\bm{r})}{Z_N}$. 
We denote the intermolecular interaction energy of the $N$ molecules as
$U_N$ and the interaction energy of these molecules with the external
field (the potential due to the presence of the solvent) as
$\psi_N$. The potential energy of the inserted $(N+1)$-st molecule,
including both the interactions with the remaining $N$ molecules and
the interaction with the external field, is denoted as
$\varphi$. Thus, the potential energies read $U_N+\psi_N$ and $\varphi
+ U_N + \psi_N=U_{N+1} + \psi_{N+1}$ in the $N$ and $(N+1)$ system, respectively. The
configuration of the $i$-th water molecule is specified by stating its
position and orientation, $\bm{s}_i=(\bm{r}_i,\Omega_i)$. Written out
explicitly, $f _{ex}(\bm{r})$ reads:
\begin{equation} 
f _{ex}(\bm{r}) = -\beta^{-1}\langle\rho(\bm{r})\rangle \ln \left \langle \int \mathrm{d}\bm{\Omega}\exp(-\beta
\varphi(\bm{r},\bm{\Omega}))\right
\rangle_N,
\label{eq2} 
\end{equation}
where $\beta=1/k_BT$ and the subscript $N$ denotes the average over
all configurations of the remaining $N$ molecules. In the case where
the system is homogeneous (denoted with index 0) Eq. (\ref{eq2}) reduces to the bulk density
multiplied by $\overline F _{ex,0}^1$, the latter term being equal to Widom's
formula for the excess chemical potential $\rho_{B}\overline F _{ex,0}^1\equiv\rho_{B}\mu_{ex}$.
In the case where the external potential, $\psi(\bm r)$, is
spherically symmetric, Eq. (\ref{eq2}) splits into an external field contribution,
$\langle\rho(\bm{r})\rangle \psi(\bm r)$, and an intrinsic term, $-
\beta^{-1}\langle\rho(\bm{r})\rangle\ln \left \langle \int \mathrm{d}\bm{\Omega}
\exp(-\beta u(\bm{r},\bm{\Omega}))\right \rangle_N$, with
$\psi=\psi_{N+1}-\psi_n$ and $u=U_{N+1}-U_N$. This separation 
establishes a connection with density
functional theory (DFT) and the Yvon-Born-Green hierarchy. Namely, it is known that the chemical potential of
a inhomogeneous system splits into the intrinsic chemical potential,
$\mu_{int}$, and the external field contribution, \textit{i.e.}
$\mu=\mu_{int}(\bm{r})+\psi(\bm{r})$. Explicitly, $\mu_{int} (\bm{r})$ is the
functional derivative of the intrinsic free energy functional
\cite{Widom2,Hansen}, $\frac{\delta \mathnormal{F_{int}}[\rho]}{\delta \langle \rho 
  (\bm{r})\rangle}=\mu_{int}(\bm{r})$. The intrinsic free energy functional
can be divided into ideal and excess parts and the latter
can be identified formally as the
excess intrinsic chemical potential, $\mu_{int}^{ex}(\bm{r})=\frac{\delta \mathnormal{F^{ex}_{int}}[\rho]}{\delta \langle \rho
  (\bm{r})\rangle}$ (see for example \cite{Widom2,Hansen}) or equivalently the singlet direct
correlation function, $c^1(\bm{r})$ in integral equation theories and the Yvon-Born-Green hierarchy
\cite{Widom2,Hansen,YBG}. 
Note that if
instead we had $\psi=\psi(\bm{r},\bm{\Omega})$ a DFT free energy functional
would read $F_{ex}=\int\mathrm{d}\bm{r}\int \mathrm{d}\bm{\Omega}
\langle\rho(\bm{r},\bm{\Omega})\rangle_{N} \overline F
_{ex}^1(\bm{r},\bm{\Omega})$. A factorization of the chemical
potential into intrinsic and external contributions as to be used in a
DFT-like approach would then be rather useless. Meanwhile a generalization
of the first member
of the YBG hierarchy can in fact be obtained formally by using the invariance
of the system under simultaneous rotation of the molecules and the
coordinate system \cite{Tarazona}.

The excess entropy density, $s_{ex}$, is obtained
from the thermodynamic identity $S_{ex}=-\frac{\partial F_{ex}}{\partial
  T}$. Taking the partial
derivative inside the integral and using the subscript $\Omega$ and a
shorthand notation for the angular integration,
$(\cdots)_{\Omega}=\int \cdots \textrm{d}\bm{\Omega}$
we obtain $S_{ex}=- \int\frac{\partial}{\partial
  T} f_{ex}(\bm{r})\mathrm{d}\bm{r}\equiv \int
s_{ex}(\bm{r})\mathrm{d}\bm{r} $ or explicitly 
\begin{IEEEeqnarray}{rCl} 
s_{ex}(\bm{r})=&\frac{1}{T}&\Big ( \beta^{-1}\langle
\rho(\bm{r})\rangle \ln \langle \exp(-\beta
\varphi(\bm{r},\bm{\Omega}))_{\Omega}\rangle_N\nonumber \\  
 &-& \ln \langle \exp(-\beta
\varphi(\bm{r},\bm{\Omega}))_{\Omega}\rangle_N \frac{\partial}{\partial
  \beta} \langle \rho(\bm{r})\rangle_N\nonumber \\  
&-& \langle
\rho(\bm{r})\rangle\frac{\partial}{\partial
  \beta} \ln \langle \exp(-\beta
\varphi(\bm{r},\bm{\Omega}))_{\Omega}\rangle_N\Big ).
\label{eq3} 
\end{IEEEeqnarray}
The first and third terms in Eq. (\ref{eq3}) represent the
contribution of \textit{intrinsic} excess entropy of single molecules,
$\overline{S}^1_{ex}(\bm{r})\equiv-\frac{\partial \overline{F}^1_{ex}(\bm{r})}{\partial T }$, and expresses the
number of degrees of freedom of single molecules at $\bm{r}$, while the second
term can be understood as an \textit{extrinsic} excess entropy density,
$-\overline{F}^1_{ex}(\bm{r})\frac{\partial \langle \rho
  (\bm{r})\rangle}{\partial T }$, and is
solely due to the fact that the system is inhomogeneous (it is
zero for a homogeneous system). Due to the indistinguishability of
molecules the local density can be written as $\langle \rho(\bm{r})\rangle = \langle \sum_{i=1}^N \delta
(\bm{r}_{i}-\bm{r})\rangle\equiv N \langle \delta
(\bm{r}_{1}-\bm{r})\rangle$, and we obtain after straightforward
    algebra that 
\begin{equation} 
\frac{\partial}{\partial \beta}\langle \rho(\bm{r})\rangle = \langle
\rho(\bm{r})\rangle \langle W_N \rangle_N - N \langle W_N\delta
(\bm{r}_{1}-\bm{r})  \rangle_N .
\label{eq4} 
\end{equation}
with $W_N=U_N+\psi_N$. Eq. (\ref{eq4}) gives a vanishing change in particle number when
integrated over the volume as it should (since we work in the
($N,V,T$) ensemble). By defining the average in the $N$-particle ensemble where molecule
$1$ is fixed at $\bm{r}$ (or equivalently, an average over all
configurations in which one of the molecules is located at $\bm{r}$) as
\begin{equation} 
\langle \mathcal{O} \rangle_{N}^{\bm{r}} = \frac{\langle \mathcal{O} \delta
(\bm{r}_{1}-\bm{r})\rangle_N}{\langle \delta
(\bm{r}_{1}-\bm{r})\rangle_N}
\equiv\frac{\langle \mathcal{O} \exp(-\beta \varphi (\bm{r},\bm{\Omega}))_{\Omega}\rangle_{N-1}}{\langle \exp(-\beta \varphi (\bm{r},\bm{\Omega}))_{\Omega}\rangle_{N-1}},
\label{eq5} 
\end{equation}
we find that the following relation holds
\begin{equation} 
\frac{\partial}{\partial \beta}\langle \rho(\bm{r})\rangle =\langle
\rho(\bm{r})\rangle ( \langle W_N \rangle_N - \langle W_N \rangle_N^{\bm{r}}), 
\label{eq6} 
\end{equation} 
Eq. (\ref{eq6}) expresses the fact that the change of local density with
temperature depends on the difference between the potential energy of
the system and the average energy of configurations where one
molecule is fixed at $\bm{r}$. Eq. (\ref{eq6}) tells us that the
local density will decrease with increasing temperature if the average energy of configurations with one
molecule fixed at $\bm{r}$ is more favorable than the potential energy
of the system. It indicates that a locally increased density with
respect to the bulk density (\textit{i.e.} the density sufficiently
far away from the source of the perturbation) is disfavored in terms
of the excess entropy as it naturally decreases the configurational volume in phase space. Elementary algebraic 
manipulations give the following expression for the intrinsic excess
entropy per molecule
\begin{equation} 
\overline{S}^{intr,1}_{ex}(\bm{r})=k_B\ln \langle \exp(-\beta
\varphi(\bm{r},\bm{\Omega}))_{\Omega}\rangle_N+\frac{\langle W
  \rangle_{N+1}^{\bm{r}} - \langle W \rangle_{N}}{T}
\label{eq7} 
\end{equation}
where we have dropped the index for the potential energy indicating
the number of molecules involved as it is equal to the one denoting
the ensemble average.
Using Eqns.(\ref{eq6}) and (\ref{eq7}) Eq. (\ref{eq3}) can be
rewritten in the final form
\begin{IEEEeqnarray}{rCl} 
s_{ex}(\bm{r})=&&\frac{1}{T}\langle
\rho(\bm{r})\rangle \Big (\ln \langle \exp(-\beta
\varphi(\bm{r},\bm{\Omega}))_{\Omega}\rangle_{N} \times \nonumber \\ 
&&\big \{ \beta
^{-1} + \langle W
\rangle_{N}^{\bm{r}}
-\langle W \rangle_{N} \big \} + \nonumber \\ 
&&\langle W
\rangle_{N+1}^{\bm{r}} - 
\langle W \rangle_{N}
\Big ).
\label{eq8} 
\end{IEEEeqnarray}   
Here, using $E=F+TS$, the expression for 
the excess energy density,
$e_{ex}(\bm{r})$, can be trivially written as:
\begin{IEEEeqnarray}{rCl} 
e_{ex}(\bm{r})=\langle
\rho(\bm{r})\rangle \Big (&&\ln \langle \exp(-\beta
\varphi(\bm{r},\bm{\Omega}))_{\Omega}\rangle_{N} 
\big \{
\langle W \rangle_{N}^{\bm{r}}-
\langle W \rangle_{N}  \big \} \nonumber \\  + 
&& \langle W
\rangle_{N+1}^{\bm{r}} - 
\langle W \rangle_{N}
\Big ).
\label{eq9} 
\end{IEEEeqnarray} 
Eqns. (\ref{eq2}), (\ref{eq8}) and (\ref{eq9}) represent the
fundamental local equations for the canonical ensemble. We see that
the nominally extensive property exactly generalizes to the
locally extensive property for inhomogeneous systems. The local
equations contain an intrinsic ('per single molecule') part and
an additional extrinsic contribution, $\langle \rho (\bm{r})\rangle \ln \langle \exp(-\beta
\varpi(\bm{r},\bm{\Omega}))_{\Omega}\rangle_{N}\frac{\langle W
\rangle_{N}^{\bm{r}}
-\langle W \rangle_{N}}{T}$, due to the inhomogeneity of the
system. 
The extrinsic component expresses the energy \textit{redistribution}.
It is caused by a local density change with respect to a homogeneous system due to the redistribution of
molecular positions at given local
density of energy states per molecule.
The intrinsic part reflects the local
density of energy states per molecule and the change of potential
energy of the system when a molecule is added to the ensemble at the
specified position. We deliberately avoid defining the local $pV$
term, as the scale-independent definition of the local work is \textit{ad
hoc} not possible because of its inherent non-local nature. However,
we show in the following that the
appropriate local $pV$ analogue arises naturally in the ($N,p,T$) ensemble. 

Analogous to the $(N,V,T)$ ensemble we again assume that the Gibbs
free energy density is locally extensive, $g_{ex}(\bm{r}) \equiv \overline G _{ex}^1
  (\bm{r})  \langle\rho(\bm{r})\rangle$, where $\overline G
_{ex}^1(\bm{r})$ is the excess Gibbs free
energy per molecule located in the differential volume element and
$\langle\rho(\bm{r})\rangle=\langle \sum_{i=1}^N \delta(\bm{r} -
\bm{r}_i)\rangle$ is the local density (at constant $N,p$ and
$T$) in the presence of an
external field $\psi$. The total Gibbs free energy
per molecule located in the differential volume element centered at
$\bm{r}$ is defined in terms of the configurational isothermal-isobaric partition
functions of the $N$ and $N+1$ systems: $\overline
G^1_c(\bm{r})=-k_BT\ln\frac{\Delta_{N+1}(\bm{r})}{\Delta_N}$ \cite{Frenkel}. The excess Gibbs free energy density
can be written as,
\begin{equation}
g_{ex}(\bm{r})=-\beta^{-1}\langle \rho (\bm{r}) \rangle\ln \frac{\left \langle V \int \mathrm{d}\bm{\Omega}\exp(-\beta
\varphi(\bm{r},\bm{\Omega}))\right\rangle_{N}}{\langle V \rangle_{N}},
\label{eq10}
\end{equation}   
In the limit of small
volume fluctuations, $\Delta V/V \to 0$, the volume terms cancel each other and
Eq.(\ref{eq10}) reduces to Eq.(\ref{eq2}). The fluctuating volume term
will be important
where volume fluctuations are expected to be large. Important
examples include dewetting phenomena of meso- and macro-scopic surfaces
\cite{dewet,weeks}, hydrophobicity on the mesoscale \cite{hydroph}
and dewetting transitions in nanochannels \cite{channel}. 

The factorization of $g_{ex}(\bm{r})$ into
excess enthalpy density, $h_{ex}(\bm{r})$, and $s_{ex}(\bm{r})$ can be achieved by introducing the
average volume $\langle V
\rangle_{N+1}^{\bm{r}}$ in a $(N+1)$-particle ensemble in which the $(N+1)$-st molecule position is fixed at
$\bm{r}$ using Eq. (\ref{eq5}):
\begin{IEEEeqnarray}{rCl}
g_{ex}(\bm{r}) = &-&\beta^{-1}\langle \rho(\bm{r})\rangle \ln \langle \exp(-\beta
  \varphi (\bm{r}))_{\Omega}\rangle_{N} \nonumber \\
&-&\beta^{-1}\langle \rho(\bm{r})\rangle \ln \frac{\langle V
  \rangle_{N+1}^{\bm{r}}}{\langle V \rangle_{N}}, 
\label{eq11}
\end{IEEEeqnarray}
where the first term is recognized as the (Helmholtz) free energy
density and the second term reflects the relative response of the
system volume to the
insertion of an additional molecule at a point $\bm{r}$ given fixed
$T$ and $p$. Comparing Eq. (\ref{eq11}) with the relation between the
Gibbs and Helmholtz free energies we recognize that the second term
in  Eq. (\ref{eq11}) is the exact local analogue of the $pV$ term and
immediately demonstrates the fact that the work is intrinsically performed
non-locally. At the same time it also demonstrates that the total work has a
well defined local contribution, which is exactly locally extensive. 
$h_{ex}(\bm{r})$ is obtained directly as
$g_{ex}(\bm{r})+Ts_{ex}(\bm{r})$ and all three together constitute the
exact fundamental local relation for the isothermal-isobaric ensemble.

Meanwhile, the
chemical potential is strictly constant throughout the system and can
only artificially be given a local physical interpretation, $\mu_{ex}=\int\mu_{ex}^{loc}(\bm{r}) \mathrm{d}\bm{r}$, in terms of an average
contribution per molecule $\mu_{ex}=N^{-1}\int g_{ex}(\bm{r}) \mathrm{d}\bm{r}$.    

The locally-extensive formulation of
statistical thermodynamics reduces to the standard description in case of
a homogeneous system. On the other hand it must also reduce to
the classical bulk thermodynamic formulation upon spatial integration. By noting that the Helmholtz
free
energy is related to $\mu$ via $F=N\mu-pV\equiv
N\phi$, where $\phi$ is the (Helmholtz) free energy per
molecule. 
Recalling that the ideal gas contribution to $F$ in the inhomogeneous local picture is $F_{id}=\beta^{-1}\int \langle\rho(\bm{r})\rangle
(\ln\langle\rho(\bm{r})\rangle-1)\mathrm{d}\bm{r}$ we obtain for $F=F_{id}+F_{ex}$,
disregarding kinetic terms, $\beta^{-1}\int \langle\rho(\bm{r})\rangle
\ln[\langle\rho(\bm{r})\rangle/
\langle\exp(-\beta\varphi(\bm{r,\Omega}))\rangle_{N}]\mathrm{d}\bm{r}$. According to
the potential distribution theorem \cite{Widom1} the logarithmic term is
constant, leading to  $F=N \phi$, as expected. However, this suggests
that Widom's definition of the chemical potential for an inhomogeneous
system (within the framework of the potential distribution
theory)\cite{Widom1} is in fact the free energy per particle. Clearly, for $N\to\infty$, $\mu$ and $\phi$ become equivalent.

As a case study we present an exact derivation of the surface tension of
a single component liquid droplet in absence of external fields such that the
liquid/gas intreface is on average spherical. Addressing this
problem in the framework of statistical mechanics using either
mechanical or point thermodynamical approaches is known to introduce a
degree of arbitrariness into the problem 
such that the surface of
tension (the surface in which the tension acts)
is said to be ill-defined \cite{Widom2}. For notational convenience we
assume that intermolecular interactions are spherically symmetrical
but may include many-body contributions of arbitrary order. The system
contains internal liquid ($l$) and external gaseous ($g$) bulk
phases separated by a non-uniform interfacial region ($s$). Choosing
the radius of the dividing surface, $r_s$, determines the volumes of
the bulk phases $V=V_l+V_g$. Constancy of total number of molecules
demands $N=N_l+N_g+N_s=4\pi\int r^2 \langle \rho (r) \rangle
\mathrm{d}r$. Only excess quantities contribute to the surface tension
and $F_{ex}$ of the system can be written in terms of bulk and surface
contributions $F=F_l+F_g+F_s$. Depending on $r_s$ the values of $N_s$
and $F_s$ can formally be positive or negative. By definition the surface tension, $\sigma$, represents the
excess free energy of the dividing surface per unit of its area:
$\sigma = (4\pi r_s^2)^{-1}(F-(F_g+F_l))$. Extensivity, $F_{l(g)}=N_{l(g)}\overline F
_{ex,0}^{1;l(g)}$ (where the subscript $0$ denotes the homogeneous
$N$-molecule ensemble of bulk phase $l(g)$  at the same $T$ and $V$), 
allows us to write 
\begin{IEEEeqnarray}{rCl}
\sigma = r_s^{-2}\Big\{ &\int_0^{r_s}& r^2\langle \rho (r) \rangle \left(
\overline F_{ex}^{1}(r) - \overline F_{ex,0}^{1;l} \right ) \mathrm{d}r
+ \nonumber \\
&\int_{r_s}^{\infty}& r^2\langle \rho (r) \rangle \left(
\overline F_{ex}^{1}(r) - \overline F_{ex,0}^{1;g} \right ) \mathrm{d}r
\Big \}.
\label{eq12}
\end{IEEEeqnarray}
We have $\sigma=\sigma(r_s)$ and it is known that at the actual surface of
tension the formal derivative vanishes at $r_t$, $d\sigma/dr_s|_{r_s=r_t}=0$
\cite{Widom2}. Applying the chain rule and making use of the second
fundamental theorem of calculus we can perform the derivation and
obtain an implicit equation for $r_t$:
\begin{IEEEeqnarray}{rCl}    
\frac{r_t^{3}}{2}\langle \rho (r_t)\rangle (\overline F _{ex}^{1;g} -
\overline F _{ex}^{1;l})
&=& \nonumber \\
\int_0^{r_t} r^2\langle \rho (r) \rangle \vartheta^{l}(r)\mathrm{d}r &+&
\int_{r_t}^{\infty} r^2\langle \rho (r) \rangle \vartheta^{g}(r) 
\mathrm{d}r,
\label{eq13}
\end{IEEEeqnarray}
where $\vartheta^{l(g)}(r)=\ln\frac{\exp(-\beta \overline F_{ex,0}^{1;l(g)})}{\langle \exp(-\beta
\varphi(r)) \rangle_N}$.
Once we know $\langle \rho (r) \rangle$, $\overline F_{ex,0}^{1;l(g)}$ and $\langle \exp(-\beta
\varphi(r))\rangle_N$, all of which can be readily calculated from Monte Carlo or
Molecular Dynamics simulations, we can compute the surface tension as
\begin{equation}
\sigma = \int_0^{r_t}\frac{r^2}{r_t^2} \langle \rho (r) \rangle \left
\{ \vartheta^{l}(r)H(r_t-r) + \vartheta^{g}(r)H(r-r_t) \right \}\mathrm{d}r,   
\label{eq14}
\end{equation}
where $H(x)$ denotes the Heaviside function. Analogous equations are
obtained in the case of a flat interface, where the surface of tension
is chosen such as to satisfy $N_s=0$. Eqs. (\ref{eq12}) and
(\ref{eq14}) show that the {\it local density of surface free energy},
(local density of $\sigma A$), is also exactly locally extensive. The above derivation
can be viewed as the exact accomplishment of the van der Waals original approximate
program for flat interfaces based on point-thermodynamics \cite{Widom2,Row}.  
The present method is free of arbitrariness and avoids the use of
local pressure tensors, which are known to be non-uniquely defined \cite{Widom2}. 

In this letter we propose an exact general formulation of local
thermodynamics of inhomogeneous fluids in the canonical and
isothermal-isobaric ensembles, which is valid on all length
scales without limitations with respect to nature intermolecular
interactions. By introducing the notion of a \textit{locally extensive
  quantity}
we have shown that the thermodynamic potentials, which
are inherently non-local, all have a well
defined local physical meaning. Our results show that there are two
distinct physical consequences of inhomogeneity, an \textit{intrinsic}
one due to the specific properties of the local environment giving rise
to a specific local density of states of a molecule, and an
\textit{extrinsic} one, which is explicitly due to the inhomogeneous
distribution of molecular positions throughout the system. 
The theory was used to derive an exact expression for the surface tension of a spherical liquid droplet. Our results pose the formal basis
for a general local thermodynamic description of 
inhomogeneous fluids ranging from macroscopic interfacial phenomena,
molecular solvation to
such as encountered in modern nanofluidic devices.  




\nocite{*}

\bibliography{apssamp}

\begin{thebibliography}{00}
\bibitem{footnote} Specifically, their approach demands that the first functional
  derivative with respect to the density, $\frac{\delta
    A[\rho]}{\delta \rho(\bm{r})}$, is local and that the second functional
  derivative, $\frac{\delta^2
    A[\rho]}{\delta \rho(\bm{r})\delta \rho(\bm{r'})}$, is short range
  in $\bm{r}-\bm{r'}$.
\end{thebibliography}

\end{document}